\documentclass[aps,prl,twocolumn,superscriptaddress,showpacs,footnotebib]{revtex4-1}
\usepackage{amsmath}
\usepackage{bm}
\usepackage{color}
\usepackage{graphicx}
\usepackage{epstopdf}
\usepackage{epsfig}
\usepackage{amsfonts}
\usepackage[naturalnames]{hyperref}
\usepackage{hypcap}
\usepackage{verbatim}
\usepackage{tabularx}
\usepackage{bbm}
\usepackage{esvect}
\usepackage{subfigure}
\usepackage{slashed}
\usepackage{hyperref}
\usepackage{slashed}
\usepackage{gensymb}

\usepackage[T1]{fontenc} 

\def\bea{\begin{eqnarray}}
\def\eea{\end{eqnarray}}

\def\ba{\begin{array}}
\def\ea{\end{array}}

\hypersetup{colorlinks=true, citecolor=blue, urlcolor=blue, linkcolor=blue}
\begin{document}

\title{Stabilizing fractional Chern insulators via exchange interaction in moir\'e systems}
\author{Xiaoyang Shen}
\thanks{These authors contributed equally to the work.}

\affiliation{State Key Laboratory of Low Dimensional Quantum Physics and
Department of Physics, Tsinghua University, Beijing, 100084, China}

\author{Chonghao Wang}
\thanks{These authors contributed equally to the work.}
\affiliation{State Key Laboratory of Low Dimensional Quantum Physics and
Department of Physics, Tsinghua University, Beijing, 100084, China} 
\author{Ruiping Guo}         
\thanks{These authors contributed equally to the work.}
\affiliation{State Key Laboratory of Low Dimensional Quantum Physics and
Department of Physics, Tsinghua University, Beijing, 100084, China}
\affiliation{Institute for Advanced Study, Tsinghua University, Beijing 100084, China}
\author{Zhiming Xu}
\affiliation{State Key Laboratory of Low Dimensional Quantum Physics and
Department of Physics, Tsinghua University, Beijing, 100084, China}
\author{Wenhui Duan}
\email{duanw@tsinghua.edu.cn}
\affiliation{State Key Laboratory of Low Dimensional Quantum Physics and
Department of Physics, Tsinghua University, Beijing, 100084, China}
\affiliation{Institute for Advanced Study, Tsinghua University, Beijing 100084, China}
\affiliation{Frontier Science Center for Quantum Information, Beijing, China}
\affiliation{Beijing Academy of Quantum Information Sciences, Beijing 100193, China}
\author{Yong Xu}
\email{yongxu@mail.tsinghua.edu.cn}
\thanks{}
\affiliation{State Key Laboratory of Low Dimensional Quantum Physics and
Department of Physics, Tsinghua University, Beijing, 100084, China}
\affiliation{Frontier Science Center for Quantum Information, Beijing, China}
\affiliation{RIKEN Center for Emergent Matter Science (CEMS), Wako, Saitama 351-0198, Japan}

\date{\today} 

\begin{abstract} Recent experimental discovery of fractional Chern insulator in moir\'e Chern band in twisted transition metal dichalocogenide homobilayers has sparked intensive interest in exploring the ways of engineering band topology and correlated states in moir\'e systems. In this letter, we demonstrate that, with an additional exchange interaction induced by proximity effect, the topology and bandwidth of the moir\'e minibands of twisted $\mathrm{MoTe_2}$ homobilayers can be easily tuned. Fractional Chern insulators at -2/3 filling are found to appear at enlarged twist angles over a large range of twist angles with enhanced many-body gaps. We further discover a topological phase transition between the fractional Chern insulator, quantum anomalous Hall crystal, and charge density wave. Our results shed light on the interplay between topology and correlation physics.
\end{abstract}
\maketitle

{\it Introductions.---}
Moir\'e materials based on multilayer van de Waals materials have recently become a burgeoning area with intensive study. The dominance of many-body interaction over kinetic energy plays a pivotal role in the emergence of the remarkable exotic phases, such as correlated insulators, superconductivity, and generalized Wigner crystals \cite{cao2018unconventional,cao2018correlated,li2024mapping,regan2020mott,xie2021fractional,Xie_2019}. Among the strongly correlated electron states, the investigation of fractional Chern insulators (FCIs) holds a broad appeal due to its novel nature and potential to serve as the possible platform for quantum computation \cite{nayak2008non,kitaev2003fault}. Amounts of theoretical works \cite{2021li,2021Devakul,abouelkomasan2020,2020Repellin,2021wilhelm} have predicted that FCI finds its manifestation in moir\'e materials such as twisted bilayer graphene (TBG) and twisted bilayer transition metal dichalcogenide (TMD), which has later been witnessed in the experiments \cite{Cai2023,Park2023,LiTingxin2023PRX,JuLong2024,xie2021fractional}. However, the presence of FCIs usually requires strong conditions on both dispersion and interaction\cite{2014Roy,regnault2011fractional,2011Qi,Jackson_2015,Parameswaran_2013,2021wang}. For instance, in reminiscence of the Landau levels in the two-dimensional electron gases, the partially filled moir\'e band should be nearly flat and topological. Besides, the quantum geometric characterized by the Fubini-Study metric should also be flat. The interaction should be strong enough to induce the many-body gap between the degenerated ground states and the excited states. As a result, the existence of FCIs is usually fragile against fluctuations and hence it is not easy to reach in both theoretical and experimental attempts \cite{wang2024fractional,Cai2023,xie2021fractional}. 

Searching for the feasibility of stabilizing such an exotic phase would be intriguing and significant. From the perspective of band structure, the flatness comes from manipulating the spatially modulated moir\'e potential. Such modulation also emerges in the magnetic materials with moir\'e structures \cite{paul2023giant,2018tong}. This indicates that the proximity exchange interaction may serve as a potential ingredient in tuning the moir\'e potential and further manipulating and stabilizing the novel phase in the moir\'e material \cite{paul2023giant,2024morales}.

In this Letter, we demonstrate that the proximity magnetic exchange interaction serves as an effective way to stabilize the FCIs and manipulate the correlated phases in moir\'e materials. Taking the twisted bilayer MoTe$_2$ as an example, under the proximity exchange interaction, the flat band is supported in a large range of twist angles, and the band topology exhibits a rich structure with respect to the exchange strength and the twist angle. Utilizing the exact diagonalization (ED), we find the existence of FCI in a wide range of twist angles, together with a prominent enhancement of the many-body gap.  We further investigate the competition between FCIs and the charge density waves (CDWs) at $-\frac{2}{3}$ filling under the exchange interaction. We report exotic exchange-interaction-induced topological phase transitions between the FCI, a charge-ordered phase with intrinsically topological nature, which is recently dubbed as \textit{quantum anomalous Hall crystal }(QAHC) \cite{sheng2024quantum,song2024intertwined,2024song} and topologically trivial CDW \cite{reddy2023toward,sharma2024topological,Hu_2023}.

{\it Continuum model.---}
The starting point is the single particle moir\'e Hamiltonian of the twisted bilayer MoTe$_2$ with the proximity exchange interaction \cite{Bistritzer_2011,wang2024fractional,wu2019topological,2024jia}.
\begin{equation}
    \mathcal{H}(\boldsymbol{r}) = \mathcal{H}_{\text{0}}(\boldsymbol{r})+\mathcal{H}_{\text{ex}}(\boldsymbol{r})
\end{equation}
Without exchange interaction, $\mathcal{H}_{\text{0}}$ is the Hamiltonian without the exchange interaction and can be split into two time-reversal counterparts $\mathcal{H}_{\tau = \pm1}$ with opposite valleys and spins.

\begin{equation}
\begin{aligned}
\mathcal{H}_{\tau}(\boldsymbol{r}) & =\left(\begin{array}{cc}
-\frac{\hbar^2\left(\boldsymbol{k} -\tau\boldsymbol{\kappa}_{+}\right)^2}{2 m^*}+\Delta_{\mathfrak{b}}(\boldsymbol{r}) & \Delta_T(\boldsymbol{r}) \\
\Delta_T^{\dagger}(\boldsymbol{r}) & -\frac{\hbar^2\left(\boldsymbol{k}-\tau\boldsymbol{\kappa}_{-}\right)^2}{2 m^*}+\Delta_{\mathfrak{t}}(\boldsymbol{r})
\end{array}\right),\\
\end{aligned}
\end{equation}
\begin{widetext}

\begin{figure}[t]
    \centering
\includegraphics[width = 0.98\textwidth]{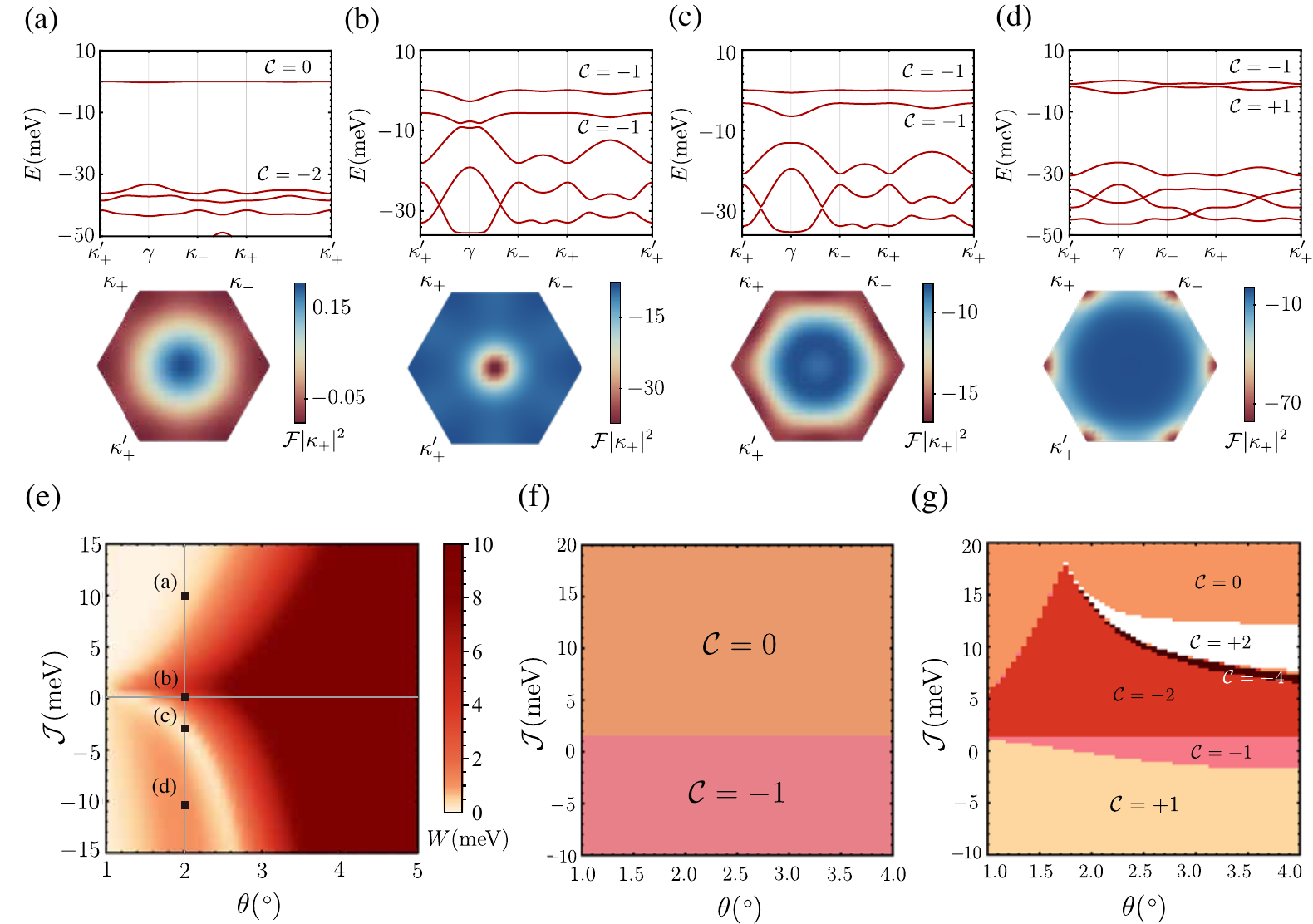}
\caption{(a)-(d) Band structure and the Berry curvature $\mathcal{F}$ at $\theta = 2^\circ, \mathcal{J} = 10,0,-2,10~\text{meV}$, respectively. (e) Bandwidth $W$ of the first valence band as a function of twist angle $\theta$ and exchange interaction strength $\mathcal{J}$. (f) Chern number $\mathcal{C}$ of the first valence band as a function of $\theta$ and $\mathcal{J}$. (g) Chern number $\mathcal{C}$ of the second valence band as a function of $\theta$ and $\mathcal{J}$. }
    \label{fig:bandwidth_chernnumber}
\end{figure}   
\end{widetext}
where
\begin{equation}
    \begin{aligned}
        \Delta_{\mathfrak{t},\mathfrak{b}}(\boldsymbol{r})& =2 V \sum_{j=1,3,5} \cos \left(\boldsymbol{g}_j \cdot \boldsymbol{r} +\ell \psi\right)\\
        \Delta_T(\boldsymbol{r})&  =  w\left(1+e^{-i \tau \boldsymbol{g}_2 \cdot \boldsymbol{r}}+e^{-i \tau \boldsymbol{g}_3 \cdot \boldsymbol{r}}\right)
    \end{aligned}
\end{equation}

$\Delta_{\mathfrak{t},\mathfrak{b}}(\boldsymbol{r})$ is the intralayer moir\'e potential, $\ell = 1$ for the top layer and $-1$ bottom layer, and $V,\psi$  capture the amplitude and shape of the potential. 
$\Delta_{T}(\boldsymbol{r})$ is the interlayer tunneling terms, and $w$ characterizes the amplitude of the tunneling. $m^*$ is the effective mass. In our numerical calculations, we take the parameters from fitting the first-principles calculation as in \cite{wu2019topological}: $V=8.0~\text{meV}$, $\psi=-89.6\degree$ and $w=-8.5~\text{meV}$. $\boldsymbol{g}_j, j = 1,\cdots,6$ are the reciprocal lattice vectors.  

$\mathcal{H}_\pm$ are related to each other via the time-reversal symmetry (TRS) operator $\mathcal{T}$. The exchange interaction $J\hat{S}(\boldsymbol{r})\cdot \hat{\sigma}$ acts on the spin which is locked to valley. Due to the dominated splitting from spin-orbit coupling (SOC) $\sim 164~\text{meV}$, the off-diagonal element $JS_x(\boldsymbol{r})\pm iJS_y(\boldsymbol{r})$ can be neglected. The position-dependent proximity exchange interaction from the proximal magnet for each $\tau$ now is
\begin{equation}
\begin{aligned}
\mathcal{H}_{\text{ex},\tau}(\boldsymbol{r}) & =\left(\begin{array}{cc}
\tau J_\mathfrak{b}S_z(\boldsymbol{r}) & 0 \\
0 & \tau J_\mathfrak{t}S_z(\boldsymbol{r})
\end{array}\right),\\
\end{aligned}
\end{equation}

 $J_\mathfrak{b(t)}$ is the strength of the exchange interaction for the bottom(top) layer and $S_z(\boldsymbol{r})$ is the position-dependent spin texture varying periodically in real space. The spin texture from the proximal magnet serves as the external source to tune the moir\'e potential. Several proposals for moir\'e magnet with highly-tunable period and giant exchange strength have been proposed and analyzed \cite{2023xie,paul2023giant,2018tong}. Without loss of generality, the simplest $C_6$ symmetric spin texture commensurate with the underlying moir\'e lattice, together with the layer-independent interaction strength $J_\mathfrak{b,t} = J$ are assumed in the following. The exchange interaction strength from the averaged magnetization in the moir\'e unit cell (MUC) and the first Fourier component is denoted  as $\mathcal{J}_0$ and $\mathcal{J}$
 \begin{widetext}

\begin{figure}[t]
    \centering
    \includegraphics[width=0.98\linewidth]{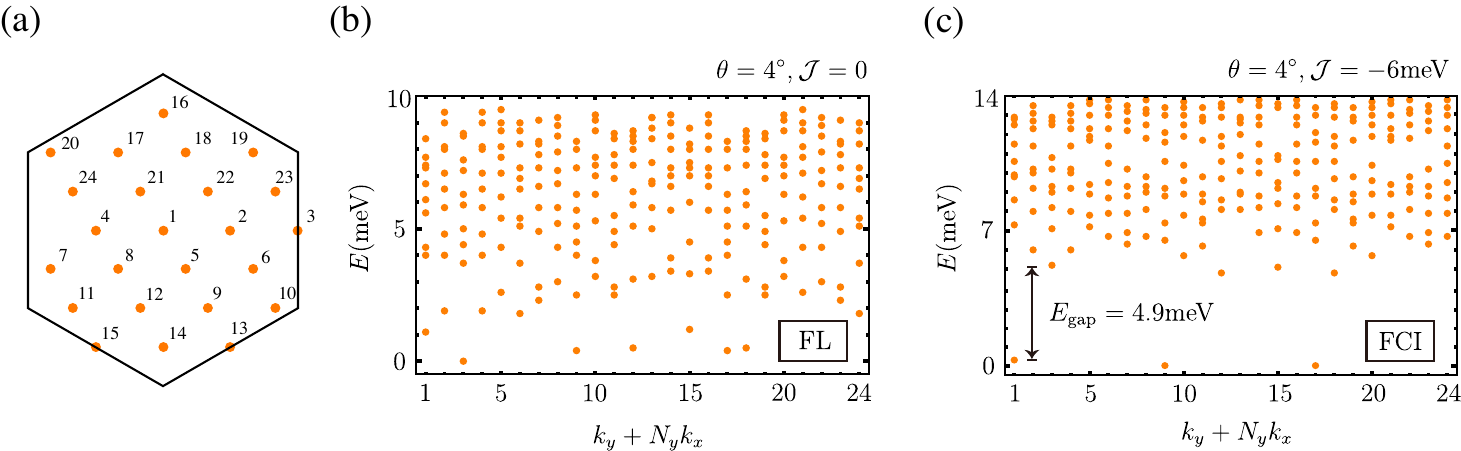}
    \caption{(a) Distribution of $4\times 6$ k-point grid in ED. (b)(c) Many-body spectrum distribution as functions of center-of-mass momentum at $\theta = 4^\circ$ with $\mathcal{J} = 0$ and $\mathcal{J} = -6~\text{meV}$, respectively. The Fermi liquid (FL) for vanishing exchange interaction strength $\mathcal{J} = 0$ and FCI with a sizable many-body gap $(\sim 4.9~\text{meV})$ for $\mathcal{J} = -6~\text{meV}$ are observed.}
    \label{ED_gap}
\end{figure}

\end{widetext}
\begin{equation}
    \begin{aligned}
        \mathcal{J}_0  = J\int_{\mathrm{MUC}} \mathrm{d}^2\boldsymbol{r} S_z(\boldsymbol{r}),\quad \mathcal{J}  = J\int_{\mathrm{MUC}} \mathrm{d}^2\boldsymbol{r} \mathrm{e}^{i \boldsymbol{g}_j \boldsymbol{r}}S_z(\boldsymbol{r}).\\
    \end{aligned}
\end{equation}
and short-wavelength components are neglected. The exchange interaction breaks $\mathcal{T}$, lifting the degeneracy between different valleys, rendering the system a valley-polarized Chern insulator at integer filling.

{\it Band structure and topology.---}
The constant Zeeman energy $\mathcal{J}_0$ does not change the band structure and topology and only shifts the relative position of valence bands from different valleys. We now fix $\mathcal{J}_0$ to a large value to guarantee the topmost valence bands are from a single copy of valley. $\mathcal{J}$ plays a central role in tuning the bandwidth and endowing a rich structure of band topology, as illustrated in Fig.\ref{fig:bandwidth_chernnumber}. Fig.\hyperref[fig:bandwidth_chernnumber]{1 (e)} illustrates the bandwidth as the function of $\theta$ and $\mathcal{J}$. The band is flattened in the presence of exchange interaction, and the flat band exists in a wide range of twist angles.  The magic angle where the ideal flat band emerges increases with the absolute value of $\mathcal{J}$.

The extensiveness of the nearly zero bandwidth implies that the exchange interaction may help promote the robustness of the flat bands. Fig.\hyperref[fig:bandwidth_chernnumber]{1 (f)(g)} exhibits the Chern numbers of the first and second valence bands. Fig.\hyperref[fig:bandwidth_chernnumber]{1 (f)} depicts a topological phase transition induced by the exchange interaction at the critical value of $ 1.56~\text{meV}$.
When $\mathcal{J}$ exceeds $1.56~\text{meV}$, the Chern number of the second band also changes from $\mathcal{C} = -1$ to $\mathcal{C} = -2$ which is attributed to the gap closing at $\gamma$ point between the first and the second valence band. The Chern number of the first valence band keeps vanishing and the bandgap increases with $\mathcal{J}$. However, the second and the third bands get closer and interact with each other with frequent band inversions, leading to a rich topological phase diagram in Fig.\hyperref[fig:bandwidth_chernnumber]{1 (g)}.

To gain a deeper view of the topology and the origin of the flat band, we track the band structure and Berry curvature of the first valence band along $\mathcal{J}$ with fixed angle $\theta = 2^\circ$. When $\mathcal{J} = 10~\text{meV}$ as shown in Fig.\hyperref[fig:bandwidth_chernnumber]{1 (a)}, the first valence band is perfectly flat and topologically trivial, serving as a potential candidate for hosting the stripe phase.  As $\mathcal{J}$ decreases, the topological phase transition happens, and the Berry curvature at $\mathcal{J} = 0$ is mainly concentrated at $\gamma$, as shown in Fig.\hyperref[fig:bandwidth_chernnumber]{1 (b)}. When $\mathcal{J}$ is negative, as shown in Fig.\hyperref[fig:bandwidth_chernnumber]{1 (b-d)}, the Berry curvature starts to transfer from $\gamma$ to $\kappa$ points with the Chern number $\mathcal{C} = -1$ fixed. Hence there exists an intermediate point with uniformly distributed Berry curvature and flat band structure.

It is worth noting that several sets of parameters $(V,w, \psi)$ have been proposed by fitting from different density functional theory results \cite{reddy2023fractional,xu2024maximally,wang2024fractional}. In practice, we test for different sets of parameters and get similar results, demonstrating the robustness and reliability of our conclusions.

{\it Stabilize the Fractional Chern insulator.---}
Having established the existence of isolated flat Chern bands appearing over a wide range of twist angles, we investigate whether FCIs can be stabilized by proximity-induced exchange interaction. We solve the many-body Hamiltonian using the exact diagonalization methodology. 
The Coulomb interaction is
\begin{equation}
        \mathcal{H}_{\mathrm{int}} = \frac{1}{2A} \sum_{\boldsymbol{k},\boldsymbol{k}^{\prime},\boldsymbol{q}}^{\mathrm{BZ}}   \sum_{\{n_i\}}
\lambda_{n_1,n_2,n_3,n_4}^{\boldsymbol{k},\boldsymbol{k}^\prime,\boldsymbol{q}}c^{\dagger}_{n_1,\boldsymbol{k}+\boldsymbol{q}} c^{\dagger}_{n_2,\boldsymbol{k}^{\prime}-\boldsymbol{q}} c_{n_3,\boldsymbol{k}^{\prime}} c_{n_4,\boldsymbol{k}},
\end{equation}
where $c^{\dagger}_{n,\boldsymbol{k}}(c_{n,\boldsymbol{k}})$ is the creation(annihilation) operator of the Bloch state, $A$ is the total area of the system, and ${n_1,n_2,n_3,n_4}$ are band indices. The form factor $\lambda_{n_1,n_2,n_3,n_4}^{\boldsymbol{k},\boldsymbol{k}^\prime,\boldsymbol{q}}$ is given by:
\begin{equation}
\begin{aligned}
\lambda_{n_1,n_2,n_3,n_4}^{\boldsymbol{k},\boldsymbol{k}^\prime,\boldsymbol{q}}= \sum_{\boldsymbol{G}}  V(\boldsymbol{q}+& \boldsymbol{G}) 
     \left\langle u_{n_1, \boldsymbol{k}+ \boldsymbol{q}+\boldsymbol{G}}\middle| u_{n_4,\boldsymbol{k}} \right\rangle \\
     & \times \left\langle u_{n_2, \boldsymbol{k}^{\prime}-\boldsymbol{q}-\boldsymbol{G}} \middle| u_{n_3,\boldsymbol{k}^{\prime}} \right\rangle.
\end{aligned}
\end{equation}
where $\left| u_{n,\mathbf{k}} \right\rangle$ is the periodic part of the Bloch state, $V(\boldsymbol{q}) = e^2 \mathrm{tanh}(|\boldsymbol{q}|d)/2\epsilon_0 \epsilon |\boldsymbol{q}|$ is the double-gated screened Coulomb potential with gate distance $d$, and $\epsilon$ is the dielectric constant \cite{wang2024fractional}. Since the exchange interaction explicitly breaks TRS, we are permitted to project the interaction onto the topmost valence band from a single valley. From the perspective of symmetry, TRS-breaking phases such as the Chern insulator, ferromagnet, and the FCI are preferred. In the following, the calculation is performed with $4\times 6$ mesh of grids in the first Brillouin zone. The gate distance is chosen to be $30~\text{nm}$ and the dielectric constant is set to be 6.

To investigate the existence of FCI, we start with some representative samplings. Two samplings with $\mathcal{J} = 0$ and $\mathcal{J} = -6~\text{meV}$ at fixed angle $\theta = 4^\circ$ are first analyzed. In the absence of exchange interaction, from Fig. \hyperref[ED_gap]{2 (b)}, there is neither three-fold-degeneracy of the ground states nor a sizable gap, which can be attributed to the dispersive band structure with Berry curvature mainly concentrating at $\gamma$ point. Thus the ground state is the Fermi liquid (FL) for small $\mathcal{J}$. However, as we increase the exchange interaction strength, the band is flattened and FCI starts to emerge as shown in Fig. \hyperref[ED_gap]{2 (c)} with a many-body gap $E_\text{gap} = 4.9~\text{meV}$. The many-body gap is defined as the energy difference between the first excited states and the three-fold-degenerate ground states. The three-fold-degenerate ground states respect the generalized Pauli principle and change with the period of $3\times2\pi$ under flux insertion.  

  \begin{figure}
    \centering
    \includegraphics[width = 0.48\textwidth]{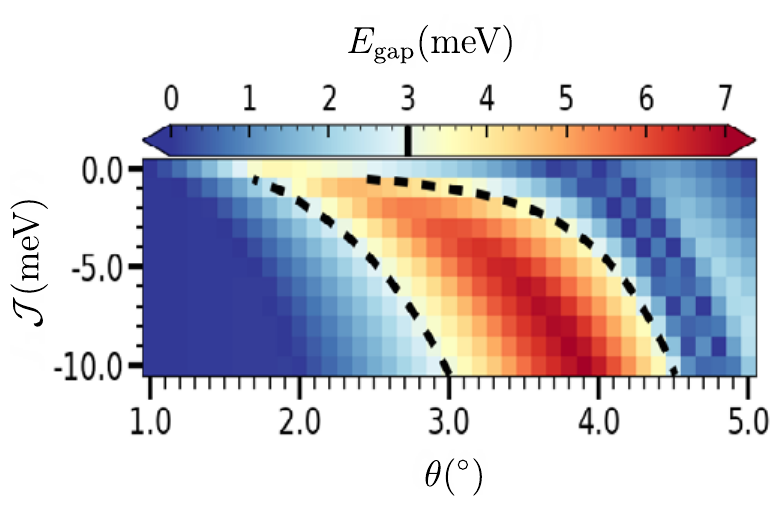}
    \caption{Many-body gap $E_{\text{gap}}$ as functions of $\mathcal{J}$ and $\theta$. $E_{\mathrm{gap}}\equiv E_4 - E_3$ where $E_i$ is the many-body eigenstate of $i^{th}$ lowest energy as a function of twist angle $\theta$ as well as $\mathcal{J}$. The dashed line corresponds to $E_{\mathrm{gap}}=3~\text{meV}$.  The ED calculation is carried out with $4\times 6$ unit cells; the gate-distance is chosen as $d=300$ \AA ~and the relative dielectric constant $\epsilon$ is 6.}
    \label{fig:many-bodygap}
\end{figure}

The many-body gap $E_\text{gap}$ is further evaluated by sweeping twist angle $\theta$ and the strength of the exchange interaction $\mathcal{J}$, as shown in Fig. \ref{fig:many-bodygap}. Remarkably, the many-body gap is prominently enhanced with the increasing strength of the exchange interaction $\mathcal{J}$. The range of the twist angle with the sizable gap grows with the exchange interaction strength, implying that under the exchange interaction, the emergence of FCI may be robust to the twist angle. The dashed line in Fig. \ref{fig:many-bodygap} corresponds to $E_{\mathrm{gap}} = 3~\text{meV}$, at which the notable enhancing effect can be observed.
The above analysis clearly demonstrates that FCIs can be stabilized by proximity-induced exchange interaction.

\textit{Exchange interaction driven phase transition.---} In the previous section, we report a topological transition in the flat band from $\mathcal{C} = 0$ to $\mathcal{C} = -1$ driven by the exchange interaction and a stabilized FCI when $\mathcal{C} = -1$.
The flat band with zero Chern number naturally serves as a platform for hosting the stripe phase, such as the generalized Wigner crystal and CDW.
It would be intriguing to unravel the intertwining between the CDW and FCI when the topologically trivial band is endowed with a non-vanishing Chern number.

\begin{figure}
    \centering
    \includegraphics[width = 0.49\textwidth]{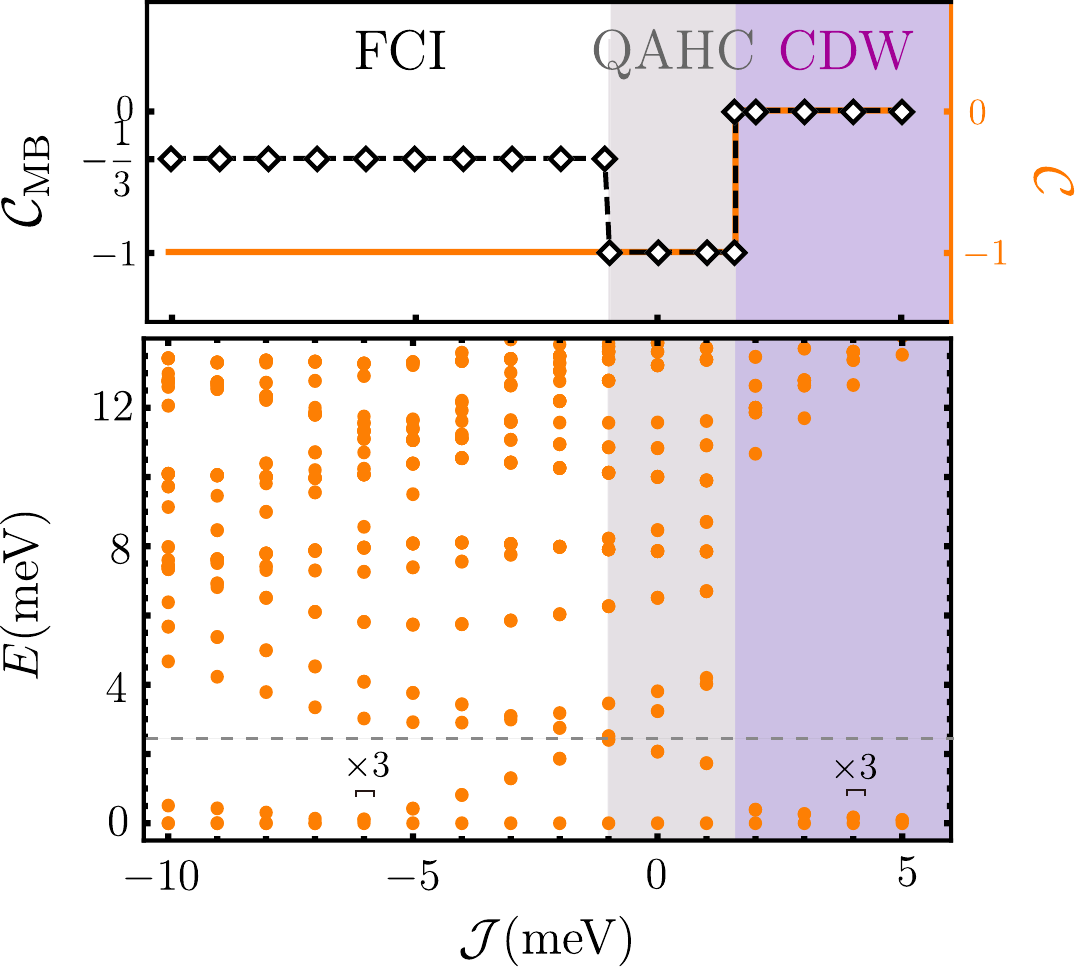}
    \caption{The upper panel is the many-body Chern number $\mathcal{C}_\text{MB}$ (black) and Chern number $\mathcal{C}$  (orange) changing with exchange interaction.  The lower panel is the spectrum changing with exchange interaction strength at $\theta = 4^\circ$. Two panels share the same horizontal axis.}
    \label{fig:QAHC}
\end{figure}

In the following, we perform an exact diagonalization calculation and track the ground states along with $\mathcal{J}$ at a fixed twist angle. The exact diagonalization is carried out on the grid of 12 k-points in BZ in the lower panel of Fig.\ref{fig:QAHC}. These 12 k-points maintain the rotation symmetry, including the high symmetry points $\gamma$, $\kappa$, $\kappa^{\prime}$, and $\rm{M}$. The dielectric constant $\epsilon$ is set to be $6$ and the twisted angle is fixed to be $4^\circ$. For each $\mathcal{J}$, we assemble the states with different center-of-mass momentum and track the first several states with the lowest energy. When $\mathcal{J}$ ranges from $-10$ to $-1~\text{meV}$, the ground state exhibits three-fold degeneracy with a sizable many-body gap increasing with the absolute value of $\mathcal{J}$. As illustrated in the previous result, the ground state obeys the generalized Pauli principle \cite{regnault2011fractional,bergholtz2008quantum} and the phase is identified as the FCI. As $\mathcal{J}$ approaches $-1~\text{meV}$, the energy gap between the third state and the fourth state decreases, and the level crossing happens at $\mathcal{J} = -1~\text{meV}$, implying some subtle change of topological order of the phase. 
When $\mathcal{J}$ exceeds $1.56~\text{meV}$, the topological phase transition from $\mathcal{C} = -1$ to $0$ takes place, rendering a discontinuous redistribution of the ground state. When $\mathcal{J}$ is positive away from 0, an enormous charge gap $\sim 10~\text{meV}$ emerges and 
we find a three-fold degeneracy at the $\kappa,\kappa^\prime$ and $\gamma$ points independent of the choice of the grids. Together with the vanishing Chern number, we identify the phase as the topologically trivial $\sqrt{3}\times \sqrt{3}$ CDW.

To further unravel the topological order of these phases, we evaluate the many-body Chern number as a probe. The many-body Chern number is defined as the integral of Berry curvature over the phase space of the twist boundary condition \cite{kudo2019many,niu1985quantized}. 
\begin{equation}
    \mathcal{C}_{\text{MB}} = \frac{i}{2\pi}\iint_0^{2\pi} \mathrm{d}\theta_x\mathrm{d}\theta_y\left(\left.\left\langle\frac{\partial\psi}{\partial\theta_y}\right|\frac{\partial\psi}{\partial\theta_x}\right\rangle-\text{c.c.}\right)
\end{equation}
where $\psi$ is the many-body wave function and $\theta_x$ and $\theta_y$ are the angles of twist boundary condition along two lattice vectors. We perform the numerical evaluation of Berry curvature under a  $10\times 10$ discretization of phase space.  
The Chern number and many-body Chern number are provided in the upper panel of Fig.(\ref{fig:QAHC}). We demonstrate two topological phase transitions as changing $\mathcal{J}$. The first phase transition takes place when $\mathcal{J} = -1\text{meV}$ between the $\nu  = -\frac{2}{3}$ FCI and the QAH with Hall conductance $\sigma_H = -1$ even when at the fractional filling. The latter phase originates from the interplay between the CDW and FCI and is recently dubbed the quantum anomalous Hall crystal (QAHC). 
The QAHC is a sort of topological charge density wave, as the phase resides at a topologically nontrivial band and exhibits the $\sqrt{3}\times \sqrt{3}$ charge order.
A way of interpreting the phase is that the emergence of $\sqrt{3}\times \sqrt{3}$ charge order spontaneously breaks the transversal symmetry, enlarging the unit cell 3 times, hence at $-\frac{2}{3}$ filling the mini-band is totally filled, leaving the integer Hall conductance $\sigma_H = -1$. The second topological phase transition between the QAHC and CDW is induced by the single-particle topology, which changes the Bloch wavefunctions discontinuously at gap closing. Observation of the Hall conductance varies from $\sigma_{H} = -1$ to $\sigma_H = 0$ is a typical signal of the phase transition.

\textit{Conclusions.---}
In this Letter, we study the twisted bilayer TMD in the presence of moir\'e proximity exchange interaction. We find that the exchange interaction assists in flattening the band structure. The wide-ranged existence of the flat band and the explicit breaking of TRS provide a platform for extensively hosting the FCIs. By carrying out the ED, we confirm the stabilizing of FCIs by observing a prominently enhanced many-body gap. Besides, the band topology also exhibits a rich structure driven by the exchange interaction. We report an exotic quantum phase transition between CDW, QAHC, and FCI induced by exchange interaction. 
In summary, we believe that our work not only leads to concrete support that proximity exchange interaction serves as an effective and practical way of manipulating the band structure and correlated electronic phase, but also sheds light on the profound interplay between topology, symmetry, and correlation.

\bibliographystyle{unsrt}
\bibliography{reference.bib}




\end{document}